\documentclass{article}
\usepackage{spconf,amsmath,graphicx}

\usepackage{times}
\usepackage{latexsym}
\usepackage{multirow}
\usepackage{makecell}
\usepackage{array}
\usepackage[T1]{fontenc}
\usepackage[utf8]{inputenc}
\usepackage{microtype}
\usepackage{graphicx}
\usepackage{mathrsfs}
\usepackage{amsmath}
\usepackage{makecell}

\usepackage{amsfonts}
\usepackage{booktabs}
\usepackage{xcolor}

\title{gated multimodal fusion with contrastive learning for turn-taking prediction in human-robot dialogue}
\name{Jiudong Yang$^1$$^{\dagger}$, Peiying Wang$^1$$^{\dagger}$\thanks{$^{\dagger}$ Equal contribution.}, Yi Zhu$^{1,2}$, Mingchao Feng$^1$, Meng Chen$^1$, Xiaodong He$^1$}
\address{
  $^1$JD AI, Beijing, China, $^2$LTL, University of Cambridge\\
  \texttt{cdyangjiudong3@jd.com, wangpeiying3@jd.com, yz568@cam.ac.uk}\\
  \texttt{fengmingchao@jd.com, chenmeng20@jd.com, xiaodong.he@jd.com}
  }
\begin{document}
%
\maketitle
\begin{abstract}
Turn-taking, aiming to decide when the next speaker can start talking, is an essential component in building human-robot spoken dialogue systems. 
Previous studies indicate that multimodal cues can facilitate this challenging task. 
However, due to the paucity of public multimodal datasets, current methods are mostly limited to either utilizing unimodal features or simplistic multimodal ensemble models. 
Besides, the inherent class imbalance in real scenario, e.g. sentence ending with short pause will be mostly regarded as the end of turn, also poses great challenge to the turn-taking decision.
In this paper, we first collect a large-scale annotated corpus for turn-taking with over 5,000 real human-robot dialogues in speech and text modalities.
Then, a novel gated multimodal fusion mechanism is devised to utilize various information seamlessly for turn-taking prediction. 
More importantly, to tackle the data imbalance issue, we design a simple yet effective data augmentation method to construct negative instances without supervision and apply contrastive learning to obtain better feature representations.
Extensive experiments are conducted and the results demonstrate the superiority and competitiveness of our model over several state-of-the-art baselines.

\end{abstract}

\section{Introduction}
\label{sec:intro}

\begin{figure}[t]
    \label{fig:example}
    \centering
    \includegraphics[width=8.5cm]{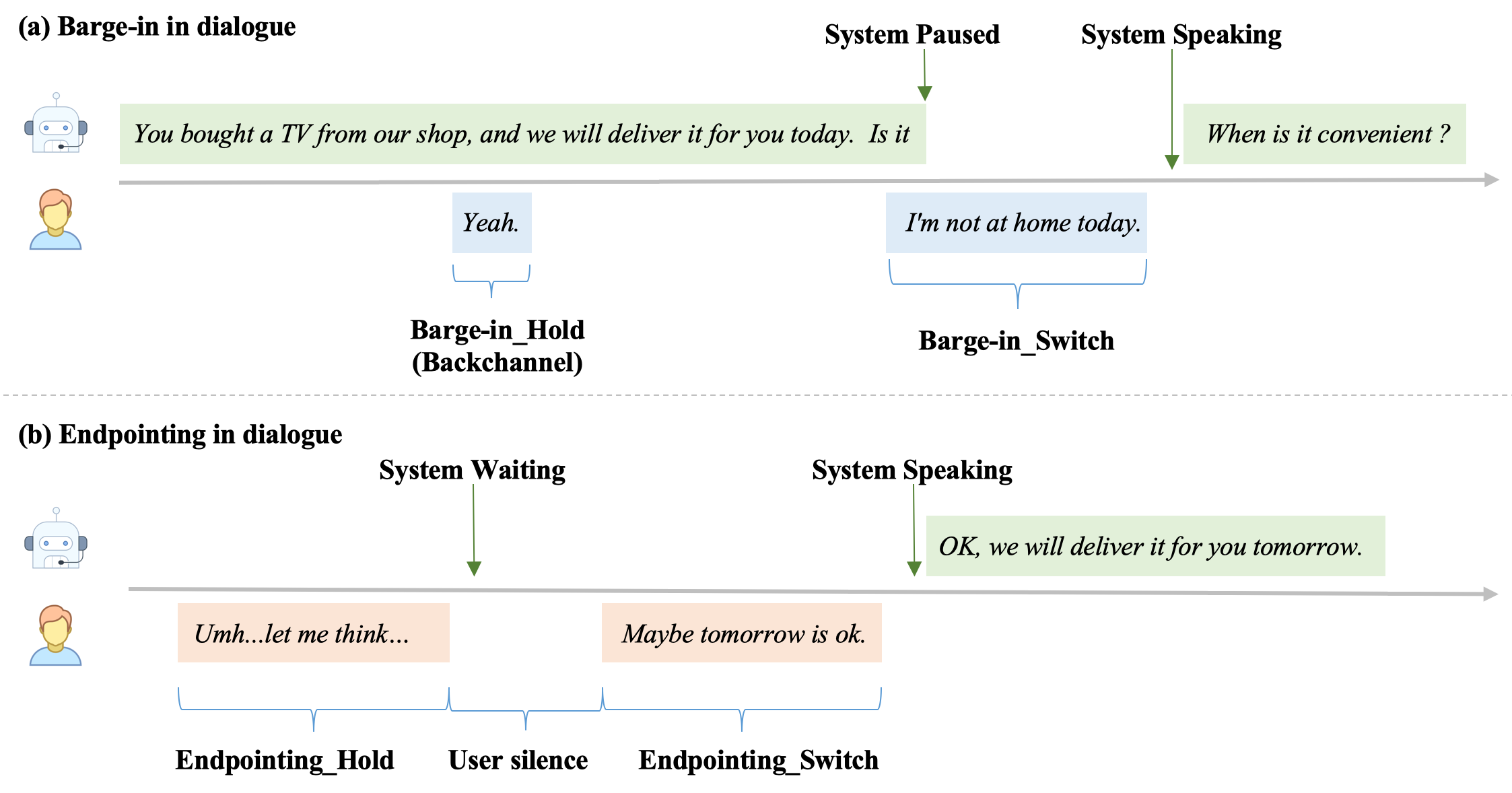}
    \caption{Example of turn-taking in a conversation.}
\vspace{-2mm}
\end{figure}

For spoken dialog systems, turn-taking is an essential component which allows participants in a dialogue to exchange control of the floor \cite{cafaro2016effects}. Given an utterance in a conversation, a \textbf{hold} means that the next utterance will be continued by the same speaker while a \textbf{switch} indicates that the next utterance will be uttered by the other speaker. 
For human-robot conversations occurred on the telephone with Interactive Voice Response (IVR) systems, turn-taking plays a critical role for user in providing natural interaction experience. 

Most of previous works in turn-taking focus on the user end-of-turn detection, i.e. \textbf{endpointing}. 
It assumes that turn switch occurs when a speaker has stopped speaking and a period of silence comes out.
Traditionally, a naive approach for endpointing is that when the current speaker pauses for a heuristically designed threshold~\cite{skantze2020turn}, the system will take the turn. 
However, this approach is limited in its naturalness that the fixed threshold can potentially be too short (frequent interruptions) or too long (awkward pauses). 
To address this problem, machine learning methods have gained popularity since 1970s~\cite[\textit{inter alia}]{duncan1972some,khouzaimi2015optimising}, and models based on inter-pausal unit (IPU), an audio segment followed by silence longer than 200 milliseconds, have mostly been studied recently because of its simplicity in practice~\cite{roddy2018investigating,liu2019neural}.
For a specific IPU, various cues across modalities, such as prosody, semantics, syntax, breathing, gesture, and eye-gaze can be extracted and integrated to determine whether this turn is yielded or not~\cite{masumura2017online,rohlfing2019multimodal}.

Although remarkable progress has been made, some issues are still present in turn-taking research. 
(1) There is a dearth of public multimodal dataset for turn-taking from real scenario: previous works mostly experiment on private in-house datasets~\cite{DBLP:conf/interspeech/HaraITK18}, pure text corpus transcribed from dialogues~\cite{roddy2018investigating,ekstedt2020turngpt}, and constructed dataset with Wizard-of-Oz setup~\cite{DBLP:conf/icassp/AldenehDP18,DBLP:conf/interspeech/ComanYM0R19} which is difficult to extract fine-grained speech information such as timing. 
Moreover, most of them ignore handling user interruptions (\textbf{barge-in}), where switch occurs when a speaker starts uttering before the other speaker finishes speaking~\cite{skantze2017towards}. Barge-in detection is crucial when the system asks longer questions or gives longer instructions which the user might have heard before or can be predicted from context. 
Figure~\ref{fig:example} shows a dialogue example of both endpointing and barge-in.
(2) Current multimodal approaches mainly use recurrent neural network (RNN) \cite{roddy2018investigating, hara2019turn} to deal with the feature sequences, whereas more advanced and efficient neural models such as Transformer \cite{vaswani2017attention} are not fully explored. 
Besides, when combining the features from different modalities, only simple ensemble techniques \cite{razavi2019investigating} are utilized, which can not optimize all feature extractors jointly. 

Motivated by above limitations, in this paper, we first collect a large-scale human-robot dialogue corpus from online conversation IVR system
in real scenario (\S\ref{sec:format}). The dataset covers both endpointing and barge-in situations, and contains more than 5,000 dialogues. 
Then we propose a novel \textbf{G}ated \textbf{M}ultimodal \textbf{F}usion model (denoted as \textbf{GMF}) for turn-taking prediction based on IPU in spoken dialogue system. GMF contains extendable feature extractors to obtain features from speech and text modalities (\S\ref{sec:pagestyle}). 
Specifically, the prevalent Transformer \cite{vaswani2017attention} and ResNet \cite{he2016deep} blocks are employed for processing text and speech respectively, and finer-grained timing features from dialogue are also considered.
Additionally, to alleviate the issue of class imbalance stemming from the characteristics of turn-taking dataset, we perform data augmentations by constructing samples for the minority class with self-supervised methods combined with constrastive learning. Extensive experiments were conducted to compare with several state-of-the-art baselines, and the results demonstrate the effectiveness of our proposed model.

\section{Dataset}
\label{sec:format}


Our dataset is collected from a commercial conversational IVR system, where conversations take place between customer and intelligent robot over the phone. 
During the call, the robot tries to make an appointment with customer for the delivery time and address of purchased goods. 
Each dialogue session lasts about 1-2 minutes with around 5-10 turns, and all turns mentioning the name of customer are removed for anonymization.
We manually transcribe all speech into text, hence both speech and text information are available. 


We extract IPUs of \textit{customer speech} from corresponding channel of IVR system. Then we group the extracted IPUs into two \textit{disjoint} subsets of endpointing and barge-in with the following heuristics: IPU which does not overlap with any robot speech is identified as endpointing, whereas for barge-in the customer interrupts while robot is speaking, i.e. customer speech starts later and overlaps with robot speech. 
For both subsets, two graduate students majoring in linguistics are instructed to annotate whether the system should \textbf{switch} or \textbf{hold} for each IPU given the whole dialogue for more accurate decision.
For endpointing, \textit{switch} means that the customer has finished his/her current speech, and the robot should take the turn, whereas \textit{hold} means that the customer has not finished and wants to continue speaking. For barge-in, \textit{switch} represents that the customer interrupted the robot by saying something meaningful and wants the robot to stop talking, while \textit{hold} means that the voice from customer might be background noise or backchannels (phatic response without significant information like \textit{yeah} and \textit{uh-huh}), and the robot should ignore it and keep speaking. 
See Figure~\ref{fig:example} for annotated turn-taking labels in each case.
The Fleiss kappa score of the annotation is 0.827, indicating substantial inter-annotator agreement. 


\begin{table}[!t]
\label{tb:statistics}
\centering
\begin{tabular}{cccc}
\hline
\multicolumn{2}{c}{\textbf{Endpointing}}&\multicolumn{2}{c}{\textbf{Barge-in}}\\
{Switch}&{Hold}&{Switch}&{Hold}\\
\hline
2451&844&1942&6312\\
74.4\%&25.6\%&23.5\%&76.5\%\\
\hline
\end{tabular}
\caption{\label{citation-guide} Statistics of our dataset.}
\vspace{-2mm}
\end{table}

The final dataset consists of 5,380 dialogues in total.
Table~\ref{tb:statistics} shows the dataset statistics in both cases.
In our scenario, as the robot starts the conversation proactively and the customer usually gives short answers (e.g., confirmation), we can see that there are more \textit{switch} instances in endpointing compared to barge-in, where most \textit{tentative interruptions} are false barge-in coming from noise or backchannel, both of which are very common in dyadic conversations via telephone. 
These observations show the complexity of our dataset, and we will mitigate the class imbalance issue via data augmentation with contrastive learning in \S\ref{sec:pagestyle}.
\section{Approach}
\label{sec:pagestyle}

\subsection{Gated Multimodal Fusion Model}

Previous studies \cite{masumura2017online,rohlfing2019multimodal} have shown that turn-taking cues across different modalities can be complementary. The combination of several cues can lead to more accurate predictions of the speaker's intentions. Inspired by this, we propose a novel model (denoted as \textbf{GMF}) to fuse various multimodal features, which is illustrated in Figure 2. Three different encoders are devised to encode text, speech, and categorical or continuous features correspondingly, which intend to catch the semantic, acoustic and timing features respectively. Then a gated multimodal fusion block is devised to fuse the above representations seamlessly. Finally, the output of the fusion layer is fed into the \textit{sigmoid} function for prediction.

\noindent \textbf{Semantic features.} Intuitively, the verbal aspect of spoken language, such as the words spoken and the semantic and pragmatic information that can be derived from those, should be very important for indicating turn shifts \cite{raux2008optimizing, razavi2019investigating}. The completion of a syntactic unit is a basic requirement for considering the turn as ``finished''. Considering the powerful ability of the Transformer block \cite{vaswani2017attention} in text representation learning, we apply it to encode both the context and current utterances: 
\begin{equation}
    \mathbf{r}^{s} = Transformer_{Encoder}(\mathbf{e})
\end{equation}
where $\textbf{e}$ is the input embedding: the sum of token, position and segment embeddings. $\mathbf{r}^{s}$ is the text representation. 
\begin{figure}[!t]
    \centering
    \includegraphics[width=8.5cm]{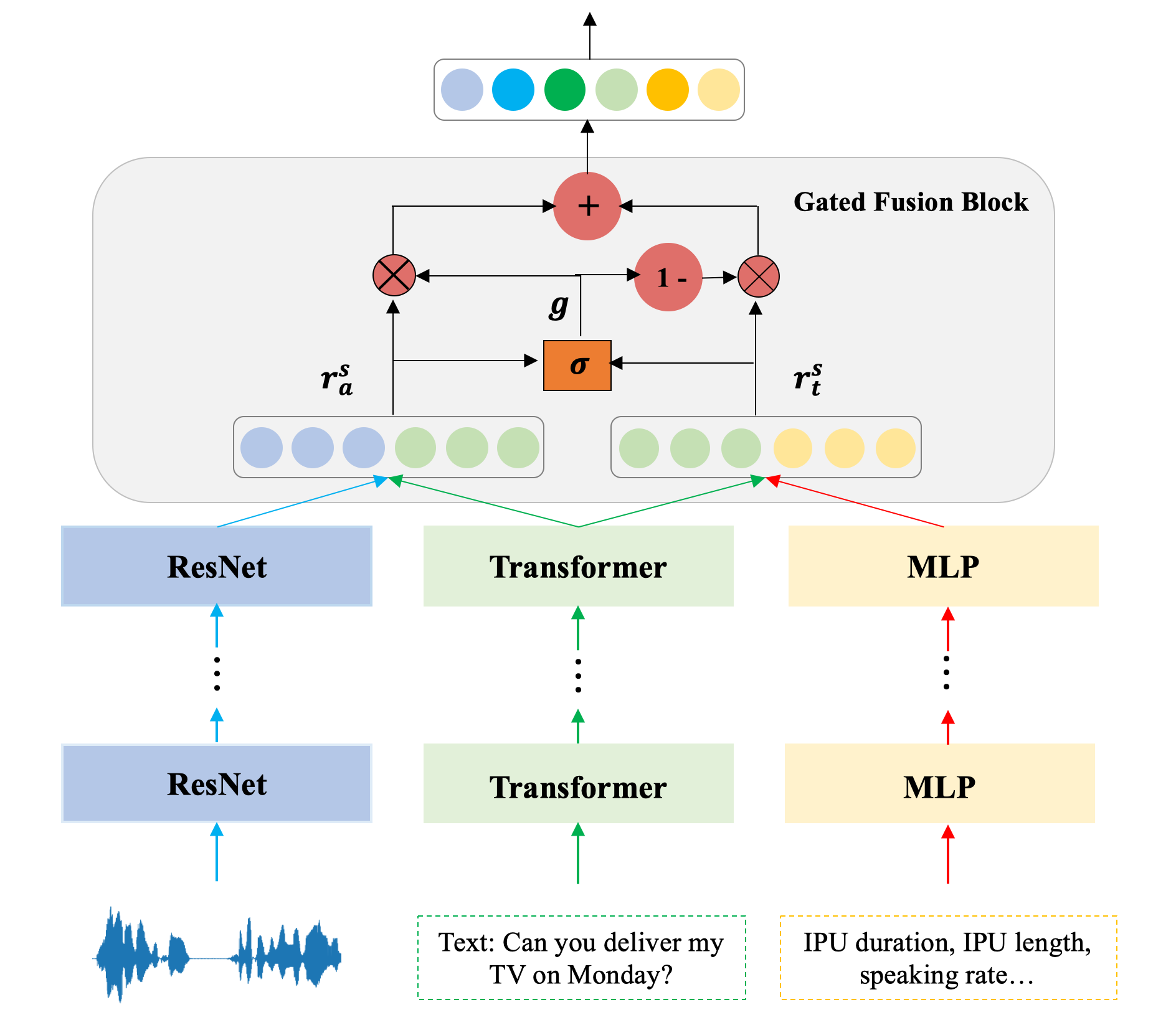}
    \caption{Architecture of our proposed model GMF.}
    \label{fig:galaxy}
\vspace{-2mm}
\end{figure}

\noindent \textbf{Acoustic features.} The role of acoustic features has been the subject of much interest in turn-taking prediction \cite{bogels2015listeners}. 
In this work, the acoustic features consist of prosodic features (e.g., \textit{energy}, \textit{pitch}), and speech features (e.g., \textit{zero-crossing-rate}, \textit{filterbank}). OpenSmile toolkit \cite{eyben2010opensmile} is used to extract above features. 
Following previous work \cite{hara2019turn}, we extract the features in the last 2 seconds of each IPU segment with a frame shift size of 50 milliseconds. For each frame, we concatenate the above mentioned features to a single vector (30 dimensions), which will be viewed as frame representation. Inspired by the successful use of ResNet architecture \cite{he2016deep} in audio tasks \cite{ford2019deep}, we feed the sequential frame representations $\textbf{f}$ into 18 ResNet layers to obtain the acoustic representation $\mathbf{r}^{a}$ as follows:
\begin{equation}
    \mathbf{r}^{a} = ResNet_{Encoder}(\mathbf{f})
\end{equation}

\noindent \textbf{Timing features.} Based on the analysis in Section 2 and previous research \cite{gravano2011turn,razavi2019investigating}, timing features can also be good indicators for turn-taking prediction. Here, we extract following four timing features, including \textit{time duration of IPU}, \textit{text length of IPU}, \textit{time interval with last turn}, and \textit{speaking rate}. All features are discretized and randomly initialized with dense vectors, then several Multilayer Perceptron (MLP) layers are applied to map timing features $\textbf{t}$ into timing representation $\mathbf{r}^{t}$:
\begin{equation}
    \mathbf{r}^{t} = MLP_{Encoder}(\mathbf{t})
\end{equation}

\noindent \textbf{Gated multimodal fusion.} After we obtain three representations from different modalities, we need to fuse them into the final representation for class prediction. Inspired by the flow control in recurrent architectures like GRU or LSTM, we devise a novel gated multimodal fusion block to control the contribution of different modalities. Considering semantic features play a vital role in turn-taking prediction, we first fuse $\mathbf{r}^{s}$ with $\mathbf{r}^{a}$ and $\mathbf{r}^{t}$ independently using fully-connected layers, resulting in $\mathbf{r}_{a}^{s} = FC(\mathbf{r}^{s}, \mathbf{r}^{a})$ and $\mathbf{r}_{t}^{s} = FC(\mathbf{r}^{s}, \mathbf{r}^{t})$. Then we combine them further as follows:
\begin{gather}
     g = \sigma(\mathbf{W}_{g}\cdot[\mathbf{r}_{a}^{s} , \mathbf{r}_{t}^{s}] ) \\
     \mathbf{r} = g\cdot\mathbf{r}_{a}^{s} +(1-g)\cdot \mathbf{r}_{t}^{s} \\
     \mathbf{\hat{y}} =\sigma(\textbf{W}_f \mathbf{r} + b) \label{con:inventoryflow}
\end{gather}
where $g$ is the gating vector, $\sigma(\cdot)$ is the \textit{sigmoid} function, and $\mathbf{\hat{y}}$ is the predicted label. $\mathbf{W}_{g}$ and $\mathbf{W}_{f}$ are weight matrices. The model is optimized by minimizing cross entropy loss $\mathcal{L}_{ce}$:
\begin{equation}
     \mathcal{L}_{ce} = -y\cdot log(\hat{y})+(1-y)\cdot log(1-\hat{y})
\end{equation}

\subsection{Data Augmentation via Contrastive Learning}
One inevitable obstacle in turn-taking prediction is the class imbalance issue. As observed and analyzed in Section 2, both endpointing and barge-in suffer from the imbalanced class distribution, which would damage the performance of classifier. To alleviate this issue, we perform data augmentations by constructing samples for the minority class with self-supervised methods and leveraging constrastive learning.

Recently, self-supervised contrastive learning (CL) has made remarkable progress in various fields~\cite{pmlr-v119-chen20j,gao2021simcse}. 
The basic idea is to pull together an anchor and a \textbf{positive} sample in embedding space, and to push apart the anchor from many \textbf{negative} samples. 
Positive examples are often obtained from data augmentations of the anchor (a.k.a views), and negative examples randomly chosen from the minibatch during training. 


Inspired by \cite{gao2021simcse}, we take dropout~\cite{DBLP:journals/jmlr/SrivastavaHKSS14} as minimal data augmentation to generate the positive pair for each sample. We randomly drop elements in the fused representation by a specific probability and set their values to zero. Besides, we also perform data augmentations for the minority class in our dataset. Specifically, for endpointing, we construct hundreds of turn-holding samples by corrupting the turn-switching samples into incomplete ones. We truncate the complete utterance (with more than 10 characters) by removing the last 30\% of words for both the speech and text. Then the utterance becomes semantically incomplete and the label is assigned as \textit{hold}. For barge-in, we first collect normal question and answer utterance pairs (i.e. system asks question and user answers the question) from dialogues. Then we move the answer utterance ahead and make it overlap with the system's question utterance in the time axis. By this way, we obtain hundreds of turn-switching samples for the barge-in scenario. Finally, we apply the contrastive loss as follows:
\begin{equation}
     \mathcal{L}_{cl} = -log\frac{e^{sim(\mathbf{x}, \mathbf{x}^{+})/\tau}}{\sum_{i=1}^{N} e^{sim(\mathbf{x}, \mathbf{x}^{-})/\tau } }
\end{equation}
where $\tau$ is a temperature hyperparameter and $sim(\cdot)$ is the cosine similarity function. As mentioned above, the positive pair is obtained by feeding the fused representation of each sample into dropout twice. The negative samples are the examples from different classes in the same minibatch, including our augmented samples with self-supervised methods.
To sum up, the total objective of our model is to minimize the following integrated loss:
\begin{equation}
     \mathcal{L} = \mathcal{L}_{ce} + \mathcal{L}_{cl}
\end{equation}

\begin{table}[!t]
\label{tb:model}
\centering
\begin{tabular}{lllll}
\hline
\textbf{Model} & \multicolumn{2}{c}{\textbf{Endpointing}} & \multicolumn{2}{c}{\textbf{Barge-in}}\\
{}&{Acc}&{Macro-F1}&{Acc}&{Macro-F1}\\
\hline
Random &\makecell[c]{0.490} &\makecell[c]{0.467} &\makecell[c]{0.512} &\makecell[c]{0.465} \\
{MajVot}$_{cls}$ & \makecell[c]{0.744} & \makecell[c]{0.425} & \makecell[c]{0.765} & \makecell[c]{0.432} \\
{LSTM}$_{ens}$ \cite{hara2019turn} & \makecell[c]{0.752} & \makecell[c]{0.646} & \makecell[c]{0.789} & \makecell[c]{0.642} \\
MoE \cite{razavi2019investigating} & \makecell[c]{0.778}  & \makecell[c]{0.643} & \makecell[c]{0.835} & \makecell[c]{0.734} \\
GMF & \makecell[c]{0.819} & \makecell[c]{0.736} & \makecell[c]{0.869} & \makecell[c]{0.814}\\
\textbf{GMF w/ CL} & \textbf{\makecell[c]{0.829}} & \textbf{\makecell[c]{0.761}} & \textbf{\makecell[c]{0.873}} & \textbf{\makecell[c]{0.826}}\\
\hline
w/o semantic & \makecell[c]{0.767} & \makecell[c]{0.658} & \makecell[c]{0.838} & \makecell[c]{0.740}\\
w/o context & \makecell[c]{0.783} & \makecell[c]{0.699} & \makecell[c]{0.852} & \makecell[c]{0.786}\\
w/o acoustic & \makecell[c]{0.788} & \makecell[c]{0.708} & \makecell[c]{0.820} & \makecell[c]{0.732}\\
w/o timing & \makecell[c]{0.791} & \makecell[c]{0.707} & \makecell[c]{0.853} & \makecell[c]{0.791}\\
\hline
\end{tabular}
\caption{Turn-taking performance of different models.}
\vspace{-2mm}
\end{table}

\section{Experiment}
\label{sec:typestyle}

\noindent \textbf{Baselines.} We compare \textbf{GMF} with the following baselines: (1) \textbf{Random}: The class is predicted randomly. 
(2) $\textbf{MajVot}_{cls}$: The class is predicted by majority voting based on class distribution of the training set. 
(3) $\textbf{LSTM}_{ens}$ \cite{hara2019turn}: It utilizes prosodic features, speech features, and linguistic features as input feature set, then three individual LSTMs are trained to catch the corresponding features, and finally a linear layer is applied to ensemble the three outputs of LSTMs. 
(4) \textbf{MoE} \cite{razavi2019investigating}: Mixture of experts that linearly interpolates four separate classifiers with SVM based on prosodic, timing, lexical \& syntactic, and semantic features. 

\noindent \textbf{Experimental Setup.} We conduct 10-fold cross validation using our dataset and report the average results. The 300-dimension Glove word embeddings are used to initialize the embedding layer of Transformer and LSTM. The number of Transformer, ResNet, and MLP layers are 3, 18, 3 respectively. The CNN kernel is set to 3 * 3 with stride of 1 in the frequency axis. The dimensions of Transformer, ResNet, and MLP are all set to 128. For all baselines, the hyper-parameters are kept consistent with the original paper. The classification \textit{accuracy} and \textit{Macro-F1} are used as evaluation metrics.

\begin{table}[t]
\label{tb:fusion} 
\centering
\begin{tabular}{lllll}
\hline
\textbf{Method} & \multicolumn{2}{c}{\textbf{Endpointing}} & \multicolumn{2}{c}{\textbf{Barge-in}}\\
{}&{Acc}&{Macro-F1}&{Acc}&{Macro-F1}\\
\hline
Concatenation & \makecell[c]{0.812} & \makecell[c]{0.723} & \makecell[c]{0.867} & \makecell[c]{0.801} \\
Summation & \makecell[c]{0.809} & \makecell[c]{0.724} & \makecell[c]{0.864} & \makecell[c]{0.799} \\
Multiplication & \makecell[c]{0.806} & \makecell[c]{0.724} & \makecell[c]{0.865} & \makecell[c]{0.801} \\
MFB \cite{yu2017multi} & \makecell[c]{0.813} & \makecell[c]{0.728} & \makecell[c]{0.861} & \makecell[c]{0.798}\\
\textbf{GMF}  &\textbf{\makecell[c]{0.819}} &\textbf{\makecell[c]{0.736}} &\textbf{\makecell[c]{0.869}} &\textbf{\makecell[c]{0.814}} \\
\hline
\end{tabular}
\caption{Turn-taking performance of different fusion methods.}
\vspace{-2mm}
\end{table}

\noindent \textbf{Main Results.} 
Table 2 shows the results on \textbf{endpointing} and \textbf{barge-in} datasets conducted separately. It's observed that, our proposed model outperforms all baselines on both datasets significantly (Sign Test, with p-value\textless0.05). Especially, \textbf{GMF} outperforms state-of-the-art approach \textbf{MoE} by absolute \textbf{9.3\%} and \textbf{8\%} on Macro-F1 score. Considering the input features are basically the same for $\textbf{LSTM}_{ens}$, \textbf{MoE} and \textbf{GMF}, it indicates that \textbf{GMF} can extract more distinguished features and fuse them more effectively. Besides, compared with $\textbf{LSTM}_{ens}$, as both Transformer and ResNet can be easily parallelized during training, \textbf{GMF} is also more efficient. As to the class imbalance issue, after we apply the data augmentation with contrastive learning (\textbf{GMF w/ CL}), it's observed that further gains up to 2.5\% Macro-F1 scores can be obtained, which demonstrates the effectiveness of contrastive learning. 

\noindent \textbf{Ablation Study.} To investigate the contribution of different components, we also conduct ablation study by removing each modality of features from GMF separately. The second part of Table 2 shows that the performance degrades correspondingly, which proves that different multimodal features are complementary to each other. We also try to remove the dialogue context from the semantic feature and use the current utterance instead, it's observed that the performance is also damaged, which illustrates the necessity of dialogue context.

\noindent \textbf{More Multimodal Fusion Approaches.} Besides the gated multimodal fusion, we also verify more multimodal fusion methods, including simple fusion methods such as concatenation ($\textbf{r} = [\mathbf{r}^{a}; \mathbf{r}^{s}; \mathbf{r}^{t}]$), summation ($\textbf{r} = \mathbf{r}^{a} + \mathbf{r}^{s} + \mathbf{r}^{t}]$), multiplication ($\textbf{r} = \mathbf{r}^{a} \circ \mathbf{r}^{s} \circ \mathbf{r}^{t}$) and more advanced information fusion approach i.e. multimodal factorized bilinear pooling (MFB)~\cite{yu2017multi}, which is widely used in visual question answering (VQA) task. Table~\ref{tb:fusion} demonstrates that the \textbf{GMF} outperforms other fusion techniques (Sign Test, with p-value\textless0.05), which indicates the advantage of gated multimodal fusion.



\section{Conclusion}
\label{sec:print}
In this paper, we focus on fusing multimodal information seamlessly to facilitate turn-taking prediction. A novel gated multimodal fusion model equipped with constrastive learning is proposed and applied on both endpointing and barge-in situations. Extensive experiments demonstrate the superiority of our model against several strong baselines. We also contribute a large-scale human-robot dialogue corpus. In the future, we will focus on exploring more turn-taking phenomena, such as backchannel and filler words. Furthermore, we will also explore more modal features (e.g., eye-gaze and gestures) to enhance our model.

\section{Acknowledgement}
\label{sec:acknowledgement}
This work is supported by the National Key R\&D Program of China under Grant No.2018YFB2100802.

\bibliographystyle{IEEEbib}
\bibliography{strings,refs}

\begin{thebibliography}{10}

\bibitem{cafaro2016effects}
Angelo Cafaro, Nadine Glas, and Catherine Pelachaud,
\newblock ``The effects of interrupting behavior on interpersonal attitude and
  engagement in dyadic interactions,''
\newblock in {\em AAMAS}, 2016.

\bibitem{skantze2020turn}
Gabriel Skantze,
\newblock ``Turn-taking in conversational systems and human-robot interaction:
  A review,''
\newblock {\em Computer Speech \& Language}, 2020.

\bibitem{duncan1972some}
Starkey Duncan,
\newblock ``Some signals and rules for taking speaking turns in
  conversations.,''
\newblock {\em Journal of personality and social psychology}, 1972.

\bibitem{khouzaimi2015optimising}
Hatim Khouzaimi, Romain Laroche, and Fabrice Lefevre,
\newblock ``Optimising turn-taking strategies with reinforcement learning,''
\newblock in {\em SIGDIAL}, 2015, pp. 315--324.

\bibitem{roddy2018investigating}
Matthew Roddy, Gabriel Skantze, and Naomi Harte,
\newblock ``Investigating speech features for continuous turn-taking prediction
  using lstms,''
\newblock {\em arXiv}, 2018.

\bibitem{liu2019neural}
Chaoran Liu, Carlos~Toshinori Ishi, and Hiroshi Ishiguro,
\newblock ``A neural turn-taking model without rnn.,''
\newblock in {\em INTERSPEECH}, 2019, pp. 4150--4154.

\bibitem{masumura2017online}
Ryo Masumura, Taichi Asami, Hirokazu Masataki, Ryo Ishii, and Ryuichiro
  Higashinaka,
\newblock ``Online end-of-turn detection from speech based on stacked
  time-asynchronous sequential networks.,''
\newblock in {\em INTERSPEECH}, 2017, pp. 1661--1665.

\bibitem{rohlfing2019multimodal}
Katharina~J Rohlfing, Giuseppe Leonardi, Iris Nomikou, Joanna
  R{\k{a}}czaszek-Leonardi, and Eyke H{\"u}llermeier,
\newblock ``Multimodal turn-taking: Motivations, methodological challenges, and
  novel approaches,''
\newblock {\em TCDS}, 2019.

\bibitem{DBLP:conf/interspeech/HaraITK18}
Kohei Hara, Koji Inoue, Katsuya Takanashi, and Tatsuya Kawahara,
\newblock ``Prediction of turn-taking using multitask learning with prediction
  of backchannels and fillers,''
\newblock in {\em Interspeech}, 2018.

\bibitem{ekstedt2020turngpt}
Erik Ekstedt and Gabriel Skantze,
\newblock ``Turngpt: a transformer-based language model for predicting
  turn-taking in spoken dialog,''
\newblock {\em arXiv}, 2020.

\bibitem{DBLP:conf/icassp/AldenehDP18}
Zakaria Aldeneh, Dimitrios Dimitriadis, and Emily~Mower Provost,
\newblock ``Improving end-of-turn detection in spoken dialogues by detecting
  speaker intentions as a secondary task,''
\newblock in {\em ICASSP}, 2018.

\bibitem{DBLP:conf/interspeech/ComanYM0R19}
Andrei~C. Coman, Koichiro Yoshino, Yukitoshi Murase, Satoshi Nakamura, and
  Giuseppe Riccardi,
\newblock ``An incremental turn-taking model for task-oriented dialog
  systems,''
\newblock in {\em INTERSPEECH}, 2019, pp. 4155--4159.

\bibitem{skantze2017towards}
Gabriel Skantze,
\newblock ``Towards a general, continuous model of turn-taking in spoken
  dialogue using lstm recurrent neural networks,''
\newblock in {\em SIGDIAL}, 2017, pp. 220--230.

\bibitem{hara2019turn}
Kohei Hara, Koji Inoue, Katsuya Takanashi, and Tatsuya Kawahara,
\newblock ``Turn-taking prediction based on detection of transition relevance
  place.,''
\newblock in {\em INTERSPEECH}, 2019.

\bibitem{vaswani2017attention}
Ashish Vaswani, Noam Shazeer, Niki Parmar, Jakob Uszkoreit, Llion Jones,
  Aidan~N. Gomez, Lukasz Kaiser, and Illia Polosukhin,
\newblock ``Attention is all you need,''
\newblock in {\em NeurIPS}, 2017.

\bibitem{razavi2019investigating}
Seyedeh~Zahra Razavi, Benjamin Kane, and Lenhart~K Schubert,
\newblock ``Investigating linguistic and semantic features for turn-taking
  prediction in open-domain human-computer conversation.,''
\newblock in {\em INTERSPEECH}, 2019.

\bibitem{he2016deep}
Kaiming He, Xiangyu Zhang, Shaoqing Ren, and Jian Sun,
\newblock ``Deep residual learning for image recognition,''
\newblock in {\em CVPR}, 2016, pp. 770--778.

\bibitem{raux2008optimizing}
Antoine Raux and Maxine Eskenazi,
\newblock ``Optimizing endpointing thresholds using dialogue features in a
  spoken dialogue system,''
\newblock in {\em SIGDIAL}, 2008, pp. 1--10.

\bibitem{bogels2015listeners}
Sara B{\"o}gels and Francisco Torreira,
\newblock ``Listeners use intonational phrase boundaries to project turn ends
  in spoken interaction,''
\newblock {\em Journal of Phonetics}, pp. 46--57, 2015.

\bibitem{eyben2010opensmile}
Florian Eyben, Martin W{\"o}llmer, and Bj{\"o}rn Schuller,
\newblock ``Opensmile: the munich versatile and fast open-source audio feature
  extractor,''
\newblock in {\em ACMMM}, 2010.

\bibitem{ford2019deep}
Logan Ford, Hao Tang, Fran{\c{c}}ois Grondin, and James~R Glass,
\newblock ``A deep residual network for large-scale acoustic scene analysis.,''
\newblock in {\em INTERSPEECH}, 2019.

\bibitem{gravano2011turn}
Agust{\'\i}n Gravano and Julia Hirschberg,
\newblock ``Turn-taking cues in task-oriented dialogue,''
\newblock {\em Computer Speech \& Language}, pp. 601--634, 2011.

\bibitem{pmlr-v119-chen20j}
Ting Chen, Simon Kornblith, Mohammad Norouzi, and Geoffrey Hinton,
\newblock ``A simple framework for contrastive learning of visual
  representations,''
\newblock in {\em ICML}, 2020.

\bibitem{gao2021simcse}
Tianyu Gao, Xingcheng Yao, and Danqi Chen,
\newblock ``Simcse: Simple contrastive learning of sentence embeddings,''
\newblock {\em arXiv}, 2021.

\bibitem{DBLP:journals/jmlr/SrivastavaHKSS14}
Nitish Srivastava, Geoffrey~E. Hinton, Alex Krizhevsky, Ilya Sutskever, and
  Ruslan Salakhutdinov,
\newblock ``Dropout: a simple way to prevent neural networks from
  overfitting,''
\newblock {\em J. Mach. Learn. Res.}, pp. 1929--1958, 2014.

\bibitem{yu2017multi}
Zhou Yu, Jun Yu, Jianping Fan, and Dacheng Tao,
\newblock ``Multi-modal factorized bilinear pooling with co-attention learning
  for visual question answering,''
\newblock in {\em ICCV}, 2017.

\end{thebibliography}

\end{document}